# Superconductivity in Topological Insulator $Sb_2Te_3$ Induced by Pressure

J. Zhu[1§], J. L. Zhang[1§], P. P. Kong[1§], S. J. Zhang[1], X. H. Yu[1,2], J. L. Zhu[1,2], Q. Q. Liu[1], X. Li[1], R. C. Yu[1], R. Ahuja[3], W. G. Yang[4], G. Y. Shen[4], H. K. Mao[4], H. M. Weng[1], X. Dai[1], Z. Fang[1], Y. S. Zhao[2,1], C. Q. Jin[1*]

1. Beijing National Laboratory for Condensed Matter Physics and Institute of Physics, Chinese Academy of Sciences, Beijing 100190, China

2. HiPSEC, Department of Physics and Astronomy, University of Nevada at Las Vegas, Las Vegas, NV 89154-4002, USA

3. Department of Physics, Uppsala University, Sweden

4. High Pressure Synergetic Consortium (HPSynC) & High Pressure Collaborative Access Team (HPCAT), Geophysical Laboratory, Carnegie Institution of Washington, Argonne, Illinois 60439, USA

**\*: Correspondence to Jin@iphy.ac.cn;**
**§: These authors contributed equally to the work.**

**Topological superconductivity is one of most fascinating properties of topological quantum matters that was theoretically proposed and can support Majorana Fermions at the edge state. Superconductivity was previously realized in a Cu-intercalated $Bi_2Se_3$ topological compound or a $Bi_2Te_3$ topological compound at high pressure. Here we report the discovery of superconductivity in the topological compound $Sb_2Te_3$ when pressure was applied. The crystal structure analysis results reveal that superconductivity at a low-pressure range occurs at the ambient phase. The Hall coefficient measurements indicate the change of p-type carriers at a low-pressure range within the ambient phase, into n-type at higher pressures, showing intimate relation to superconducting transition temperature. The first principle calculations based on experimental measurements of the crystal lattice show that $Sb_2Te_3$ retains its Dirac surface states within the low-pressure ambient phase where superconductivity was observed, which indicates a strong relationship between superconductivity and topology nature.**

**Introduction**

As new states of quantum matter, topological insulators are characterized by an insulating gap in the bulk state and a robust metallic surface or edge state protected by time-reversal symmetry (*1–4*). Topological surface states have been theoretically predicted and experimentally observed in three-dimensional materials such as $A_2B_3$-type compounds of $Bi_2Se_3$, $Bi_2Te_3$, and $Sb_2Te_3$ (*5–7*). Similar to topological insulators, topological superconductors are expected to have a full pairing gap in the bulk and gapless surface states that can support Majorana fermions at the edge states (*8–13*). Electronic excitations related to topological states, particularly Majorana fermions, are potentially useful in topological quantum computing and have thus attracted increasing attention (*14, 15*). Superconductivity in the bulk states of topological insulators together with well-defined Dirac-type surface states around the Fermi energy has been proposed to approach these novel states (*9*). Recently, superconductivity with critical transition temperature ($T_c$) at =3.8 K was observed in $Bi_2Se_3$, a typical topological insulator, by intercalating Cu between adjacent quintuple units (*16, 17*).

Apart from chemical doping, an alternative approach to induce superconductivity is to tune the electronic structure in physical manner by applying high pressure. This process possesses advantages without introducing disorders or impurities. The application of pressure has recently been reported to turn the topological insulator $Bi_2Te_3$ into a superconducting state (*18*). Isostructural to $Bi_2Te_3$ and $Bi_2Se_3$ (*19–25*), $Sb_2Te_3$ is another well-studied three-dimensional topological insulator. In this study, we report the discovery of superconductivity in $Sb_2Te_3$ single crystals induced via pressure. The nonmonotonic dependence of the superconducting transition temperature (Tc) on pressure and its relation to the Hall coefficient were observed. The experimental results, together with theoretical calculations, indicate that the superconductivity induced via pressure in the ambient phase of $Sb_2Te_3$ is related to its topological nature. Furthermore, a systematic phase diagram on crystal and electronic properties of $Sb_2Te_3$ as a function of pressure is presented.

**Results**

Fig. 1(a) shows the evolution in resistance as a function of temperature of $Sb_2Te_3$ single crystals at various pressures. Below 4.0 GPa, $Sb_2Te_3$ does not display superconductivity at temperature to 1.5 K. When the pressure was increased beyond 4.0 GPa, a superconducting transition with a $T_c$ of around 3 K was observed, as shown in Fig. 1(a). Further increasing pressure to 6.8 GPa, $T_c$ grows rapidly with the resistance drop getting more pronounced and the zero-resistance state being fully realized. A superconducting transition with higher $T_c$ was observed at 7.5 GPa, after which $T_c$ becomes constant up to 30 GPa. The pressure-induced superconductivity exhibits more complex behaviors when the pressure was further increased from 16.3 GPa to ~30 GPa. When the pressure was higher than 30 GPa, the superconducting transition becomes sharp again, which indicates the good homogeneity of a single superconducting phase with $T_c$ of about 7.3 K. We further measured the resistance versus temperature as a function of magnetic field to confirm if results shown in Fig.

1(a) are indeed superconducting transition. Fig. 1(b) exhibits the measured resistance at 6.7 GPa with applied magnetic $H$. It is obvious that the drops of resistance shift toward lower temperature with increasing magnetic field, which indicates that the transition is superconductivity in nature. The inset of Fig. 1(b) shows the magnetic $H$-dependence of $T_c$. The Werthamer-Helfand-Hohenberg formula (26) $H_{c2}(0)= -0.691[dH_{c2}(T)/dT]_{T=T_c} \cdot T_c$ was used, and the upper critical $H_{c2}(0)$ is extrapolated to be 2.6 Tesla for $H_{||c}$ when the single crystal was placed inside the diamond anvil cell with the magnetic $H$ direction perpendicular to the *ab*-plane.

The electronic properties of $Sb_2Te_3$ below 12 GPa is of particular interest because the crystal structure within this pressure range remains the same as in the ambient phase (will be discussed later) where the topological insulator behavior has been predicted and observed. Fig. 2(a) shows the pressure dependence of $T_c$ from 0 GPa to 10 GPa. The evolution of $T_c$ as a function of pressure shows an abrupt increase at around 7.5 GPa, which enables us to divide the superconducting phase diagram into three regions as follows: region A with no superconductivity, and regions B and C with superconductivity. $T_c$ in region B rapidly increases with increasing pressure at a rate of +0.45 K per GPa, whereas $T_c$ in region C slightly increases with increasing pressure at a much lower rate of +0.02 K per GPa.

We then performed Hall coefficient experiments on $Sb_2Te_3$ at high pressure. The pressure dependence of the carrier density calculated from the linear part of a high magnetic field $H$ at 2 K, 30 K and 218 K is shown in Fig. 2(b). At ambient conditions, the initial value of carrier density indicates that the $Sb_2Te_3$ as grown single crystal is of p-type carrier nature with carrier density about $5.3\times10^{19}$ /cm$^3$ similar to that of $Bi_2Te_3$ (*18*). Below 7.2 GPa, the carrier density rapidly increases with increasing pressure, and reaches about $4.7\times10^{21}$ /cm$^3$ at 7.2 GPa. After abnormal changes, the sample that is hole-dominated at ambient pressure assumes an electron-dominated character within the pressure range of 7.2 GPa–8.3 GPa, which is similar to some $Bi_2Te_3$ crystals (*25*). When the pressure was further increased, the n-type carrier density remains constant and stabilized around $8\times10^{21}$/cm$^3$.

The observed complex behavior of carrier density at high pressure invites us to study the pronounced electronic structure change hidden behind the pressure-induced superconductivity in $Sb_2Te_3$. Comparing Fig. 2(b) with Fig. 2(a), the carrier density slightly increases with increasing pressure in region A, where the sample is not superconducting. The carrier density sharply increases by almost two orders of magnitude when the pressure was further increased in region B, where superconductivity is induced. The rapid increase in the carrier density at low pressure, especially in region B, indicates that conductivity induced by pressure in the ambient phase of $Sb_2Te_3$ was significantly enhanced. When the pressure was increased to ~8 GPa, superconductivity with higher $T_c$ is induced, and accompanied the change in carrier type from hole-like into an electron-dominated superconductor. This kind of carrier type flip at high pressure was observed in several semiconductor materials, and is ascribed to the change in electronic structure, e.g., the Lifshitz phase transition (*27*).

Pressure greatly alters the electronic structure and has a pivotal function of inducing band crossing in $Sb_2Te_3$. Only a negligible increase in $T_c$ was observed when the pressure was increased in region C, in which the n-type carrier density remains almost constant at $\sim 10^{21}/cm^3$. The combined results in Fig. 2(a) and (b) strongly indicate the dependence of $T_c$ on carrier density, where a higher carrier density results in an enhanced superconducting transition temperature.

We performed *in situ* high-pressure synchrotron X-ray powder diffraction experiments on $Sb_2Te_3$ to understand the complex high-pressure behavior of $Sb_2Te_3$ comprehensively. Note that the effect of low temperature on the structural evolution of $Sb_2Te_3$ is critical for studying the electronic property of topological insulators, therefore the *in situ* high-pressure experiments, with the temperature decreased to about 8 K, were performed as shown in Fig. 3(a). The results reveal that the first phase-transition pressure is above 12.9 GPa, which indicates that the pressure-induced superconductivity observed at a pressure range of 4 GPa to 12.9 GPa indeed comes from the ambient phase. Based on the ambient pressure phase structure, the lattice parameters from 0.2 GPa to 12.9 GPa were calculated using the Rietveld refinements, as shown in Fig. 3(b). The basal lattice parameters *a* and c in rhombohedra α phase of $Sb_2Te_3$ decrease by 3.7% and 3.9% below 9 GPa, respectively. Referring to the high pressure x ray diffraction experiments for $Sb_2Te_3$ at higher pressures at room temperature reported in ref. 28, as well as the isostrutural compound $Bi_2Te_3$(*29*), four phases are assigned, i.e., α, β, γ, δ being ambient phase, the high pressure phase I, high pressure phase II, high pressure phase II, respectively(*28*). The crystal evolution information with the application of pressure in $Sb_2Te_3$ can be used to analyze the distinct pressure dependence of $T_c$. Fig. 4 illustrates the superconducting and structural phase diagram as a function of pressure up to 30 GPa.

**Discussions**

We studied both the bulk and surface states via first principle calculations by taking into account spin orbital coupling based on the experimental measurements of the crystal structure to investigate the electronic structure evolution of $Sb_2Te_3$. Firstly, we obtain the lattice parameters under different pressure by varying the volume of $Sb_2Te_3$ with fixed c/a ratio and internal atomic site. Secondly, the atomic sites are relaxed with fixed lattice parameters corresponding to specified external pressure. The theoretically optimized structure is compared with experimental value in Table 1. We present the electronic structures calculated from theoretically optimized crystal structure since those obtained from experimental structure are nearly the same. We construct the projected atomic Wannier functions (*30, 31*) for p orbitals of both Sb and Te. With this basis set, an effective model Hamiltonian for a slab of 45 QLs is established and the topologically nontrivial surface state is obtained from it. As we can see from Fig. 5, the total band gap (around Z point) of $Sb_2Te_3$ under 6.9 GPa is reduced in comparison with that under 0 GPa. However, the band gap at Γ is enlarged as pressure increases. This feature is nearly the same as in $Bi_2Te_3$ (*18*). The topologically nontrivial surface state does exist even within 6.9 GPa, although the band gap is reduced. Therefore the superconducting states observed in the high pressure phases are topological trivial as indicated in Fig.4.

However, the results show that little change was observed in the electronic structure at high pressures for the ambient phase, except for a small relative shift. A Dirac cone remains stable at a pressure of 6.9 GPa. This result provides strong support the occurrence of superconductivity in the low-pressure range of the ambient phase is topologically related. Assuming that the bulk phase becomes an S wave superconductor by applying pressure, a proximate effect changes the surface state into a superconducting one (*9*). The superconductivity at the surface state with a well-defined Dirac cone could be topological related with p+ip wave function symmetry.

**Methods**

$Sb_2Te_3$ single crystal was grown using the Bridgeman method, as described in Ref 18. Stoichiometric amounts of high purity elements Sb (99.999%) and Te (99.999%) were mixed, ground, and pressed into pills, then loaded into a quartz Bridgeman ampoule, which was then evacuated and sealed. The ampoule was placed in a furnace and heated at 800 °C for 3 days. After which, the ampoule was slowly cooled in a temperature gradient at rate 5 °C per hour to 300 °C, followed by furnace cooling. The product was cleaved easily along the basal plane. The ambient phase of the product was identified via X-ray powder diffraction.

The electronic transport properties of $Sb_2Te_3$ under high pressure at low temperatures were investigated via four-probe electrical conductivity methods in a diamond anvil cell (DAC) made of CuBe alloy (*18, 19, 32*), which has very good low-temperature properties. Pressure was generated by a pair of diamonds with a 500-μm-diameter culet. A gasket made of T301 stainless steel was covered with cubic BN fine powders to protect the electrode leads from the gasket. The electrodes were slim Au wires with a diameter of 18 μm. The gasket, preindented from a thickness of 250 μm to 60 μm, was drilled to produce a hole with a diameter of about 200 μm. The insulating layer was pressed into this hole. A 100-μm-diameter hole, which was used as the sample chamber, was drilled at the center of the insulating layer. The dimension of the $Sb_2Te_3$ single crystal was 90 μm×90 μm×10 μm, and was covered with soft hBN fine powders as a pressure transmitting medium. The pressure was measured via the ruby fluorescence method (*33*) at room temperature before and after each cooling. The diamond anvil cell was placed inside a Mag lab system upon loading. The temperature was automatically controlled by a program of the Mag Lab system. A thermometer was mounted near the diamond in the diamond anvil cell to monitor the exact sample temperature. The rate of temperature decrease was very slow to ensure the equilibrium of temperature. Hall coefficients were measured via the Van der Paul method. We investigated the pressure gradient within the sample chamber. The results show that for the geometry used in the experiments, the pressure keeps almost constant within a distance of 50 μm from the center. The gradient is less than 10% within 50 μm from the center. Our sample at the center of the chamber has a size of ~100 μm. Thus, the pressure uncertainty in our experiments is about 1 GPa at a 10 GPa scale. In situ high-pressure angle-dispersive X-ray diffraction (ADXRD) experiments were performed at a low temperature at the HPCAT of the advanced photon source (APS) with a wavelength of 0.3981 Å by using a symmetric Mao Bell DAC. An *in situ* ruby pressure calibration system was used to detect pressure at low temperature. Fine powders that were ground from the single crystal were loaded in a DAC with a tiny ruby chip as a pressure marker. The diffraction rings were recorded via image plate techniques, and the XRD patterns were integrated from the images by using the FIT2d software. A GSAS package was used to refine the crystal structures based on the Rietveld method (*34*).

The crystal structure of $Sb_2Te_3$ under different pressure is optimized within generalized gradient approximation (GGA) as parameterized by Perdew, Burke and Ernzerhof (*35*). Spin-orbit coupling is included during the optimization procedure. The first-principles calculations are performed by using OpenMX software package (*36*), which is based on linear combination of pseudo-atomic orbital (PAO) method (*37*). The PAOs are generated by a confinement potential scheme (*38*) with a cutoff radius of 9.0 and 7.5 a.u. for Sb and Te, respectively. Basis set with s2p2d2f1 PAOs for Sb and s2p2d3 for Te is found to be good enough to describe $Sb_2Te_3$.

**Authors contributions**: C.Q.J. conceived the work; J.L.Z. grown $Sb_2Te_3$ single crystals with preliminary characterizations; J.Z., P.P.K., S.J.Z, X.L., Q.Q.L., R.C.Y. conducted the high pressure transport measurements; X.H.Y., J.L.Z., W.G.Y. contributed to the measurements of high pressure structures with the helps of Y.S.Z., G.Y.S.; R.A., H.M.W., Z.F., X.D. contributed to the theoretical analysis; C.Q.J., J.Z., P.P.K. analyzed the data; C.Q.J., J.Z., wrote the paper. All authors contributed to the discussions of the work.

Additional Information: The authors declare no competing financial interests.

**References**

1. Bernevig, B. A., Hughes, T. L., Zhang, S. C. Quantum spin Hall effect and topological phase transition in HgTe quantum wells. *Science* **314**, 1757–1761 (2006).

2. Fu, L., Kane, C. L. and Mele, E. J. Topological insulators in three dimensions. *Phys. Rev. Lett.* **98**, 106803 (2007).

3. König, M., *et al.* Quantum spin Hall insulator state in HgTe quantum wells. Science **318**, 766 (2007).

4. Hsieh, D., *et al.* A topological Dirac insulator in a quantum spin Hall phase. Nature **452**, 970 (2008).

5. Zhang, H. J., *et al.* Topological insulators in $Bi_2Se_3$, $Bi_2Te_3$ and $Sb_2Te_3$ with a single Dirac cone on the surface. *Nat. Phys.* **5**, 438–442 (2009).

6. Chen, Y. L., *et al.* Experimental realization of a three-dimensional topological insulator, $Bi_2Te_3$. *Science* **325**,178–181 (2009).

7. Xia Y., *et al.* Observation of a large-gap topological-insulator class with a single Dirac cone on the surface. *Nat. Phys.* **5**, 398–402 (2009).

8. Qi, X. L., Hughes, T. L., Raghu, S., Zhang, S. C. Time-reversal-invariant topological superconductors and superfluids in two and three dimensions. *Phys. Rev. Lett.* **102**, 187001 (2009).

9. Fu, L., Kane, C. L. Superconducting proximity effect and Majorana fermions at the surface of a topological insulator. *Phys. Rev. Lett.* **100**, 096407 (2008).

10. Qi, X. L. and Zhang, S. C. Topological insulators and superconductors. Rev. Mod.

Phys. **83**, 1057 (2011).

11. Schnyder, A. P., Ryu, S., Furusaki, A., Ludwig, A. W. W. Classification of topological insulators and superconductors in three spatial dimensions. *Phys. Rev. B* **78**, 195125 (2008).

12. Ryu, S., *et al.* Topological insulators and superconductors: tenfold way and dimensional hierarchy. *New J. Phys.* **12**, 065010 (2010).

13. Akhmerov, A. R., Nilsson, J., Beenakker, C. W. J. Electrically detected interferometry of Majorana fermions in a topological insulator. *Phys. Rev. Lett.* **102**, 216404 (2009).

14. Wilczek, F. Majorana returns. *Nat. Phys.* **5**, 614–618 (2009).

15. Nayak, C., Simon, S. H., Stern, A., Freedman, M., Das Sarma, S. Non-Abelian anyons and topological quantum computation. *Rev. Mod. Phys.* **80**, 1083–1159 (2008).

16. Hor, Y. S. *et al*. Superconductivity in $Cu_xBi_2Se_3$ and its implications for pairing in the undoped topological insulator. *Phys. Rev. Lett.* 104, 057001 (2010).

17. Sasaki, S., Kriener, M., Segawa, K., Yada, K., Tanaka, Y., Sato, M., Ando, Y., Topological Superconductivity in CuxBi2Se3. *Phys. Rev. Lett.* **107**, 217001 (2011).

18. Zhang, J. L. *et al*. Pressure-induced superconductivity in topological parent compound $Bi_2Te_3$. *Proc. Natl. Acad. Sci. U.S.A.* 108, 24-28 (2011).

19. Zhang, S. J., Zhang, J. L., Yu, X. H., Zhu, J., Kong, P. P., Feng, S. M., Liu, Q. Q., Yang, L. X., Wang, X. C., Cao, L. Z., Yang, W. G., Wang, L., Mao, H. K., Zhao, Y.

S., Liu, H. Z., Dai, X., Fang, Z., Zhang, S. C. and Jin, C. Q., The comprehensive phase evolution for Bi2Te3 topological compound as function of pressure. *J. Appl. Phys.* **111**, 112630 (2012).

20. Jacobsen, M. K. *et al*. High pressure X‐ray diffraction studies of Bi2−*x*Sb*x*Te3 (x = 0,1,2). *AIP Conf. Proc.* **955**, 171 (2007).

21. Einaga, M., *et al*. New superconducting phase of Bi2Te3 under pressure above 11 GPa. *J. Phys.: Conf. Ser.* **215**, 012036 (2010).

22. Einaga, M., *et al*. Pressure-induced phase transition of Bi2Te3 to a bcc structure. *Phys. Rev. B* **83**, 092102 (2011).

23. Hamlin, J. J., *et al*. High pressure transport properties of the topological insulator Bi2Se3. *J. Phys.: Condens. Matter* **24**, 035602 (2012).

24. Vilaplana, R., *et al*. Structural and vibrational study of Bi2Se3 under high pressure. *Phys. Rev. B* **84**, 184110 (2011).

25. Zhang, C. *et al*. Phase diagram of a pressure-induced superconducting state and its relation to the Hall coefficient of Bi2Te3 single crystals. *Phys. Rev. B* **83**, 140504 (2011).

26. Werthamer, N. R., Helfand, E., Hohenberg, P. C. Temperature and purity dependence of the superconducting critical field, Hc2. III. electron spin and spin-orbit effects. *Phys Rev* **147**, 295–302 (1966).

27. Godwal, B. K. *et al*. Electronic topological and structural transition in AuIn2 under pressure. *Phys. Rev. B* **57**, 773 (1998).

28. J.G. Zhao *et al*.Pressure-Induced Disordered Substitution Alloy in Sb2Te3. *Inorg. Chem*. 50, 11291(2011).

29. Zhu, L., Wang, H., Wang, Y., Lv, J., Ma, Y., Cui, Q. & Zou, G. Substitutional alloy of Bi and Te at high pressure. *Phys. Rev. Lett.* **106**, 145501 (2011).

30. Weng, H. M., Ozaki, T. and Terakura, K., Revisiting magnetic coupling in transition-metal-benzene complexes with maximally localized Wannier functions, Phys. Rev. B **79**, 235118 (2009).

31. Zhang, W., Yu, R., Zhang, H. J., Dai, X. and Fang, Z., First-principles studies of the three-dimensional strong topological insulators $Bi_2Te_3$, $Bi_2Se_3$ and $Sb_2Te_3$, New J. Phys. **12**, 065013 (2010).

32. Zhang, S. J. *et al.* Superconductivity at 31 K in the "111"-type iron arsenide superconductor Na1-xFeAs induced by pressure. *Europhys. Lett.* **88**, 47008 (2009).

33. Mao, H. K., Xu, J., and Bell, P. M., Calibration of the ruby pressure gauge to 800 kbar under quasi-hydrostatic conditions. *J. Geophys. Res.* **91**, 4673 (1986).

34. Larson, A. C. & Von Dreele, R. B. *GSAS manual: Report LAUR* 86–748 (Los Alamos National Laboratory, (1994).

35. Perdew, J. P., Burke, K. and Ernzerhof, M., Generalized Gradient Approximation Made Simple, Phys. Rev. Lett. **77**, 3865 (1996).

36. http://www.openmx-square.org/

37. Ozaki, T. and Kino, H., Efficient projector expansion for the ab initio LCAO method, Phys. Rev. B **72**, 045121 (2005).

38. Ozaki, T. and Kino, H., Numerical atomic basis orbitals from H to Kr, Phys. Rev. B **69**, 195113 (2004).

**Acknowledgments.** The work was supported by NSF & MOST of China through research projects. W. G. Y, G. Y. S acknowledge support by EFree, an Energy Frontier Research Center funded by the U.S. Department of Energy (DOE) under Award DE-SC0001057. HPCAT is supported by CIW, CDAC, UNLV and LLNL through funding from DOE-NNSA, DOE-BES and NSF.

Table I Theoretically (theo) optimized lattice parameters of $Sb_2Te_3$ under different pressure 0, and 6.9 GPa within GGA+SOC calculation, in comparison with experimentally (exp) determined values under 6.9 GPa.

|       | 0GPa         | 6.9GPa    |         |
|-------|--------------|-----------|---------|
|       |              | theo.     | exp.    |
| a     | 4.322536333  | 4.16325   | 4.1143  |
| c     | 30.8679419028| 29.73047  | 29.1700 |
| u(Te) | 0.213299533  | 0.208001  | 0.20815 |
| v(Sb) | 0.397758538  | 0.399645  | 0.40161 |

**Figure Captions**

FIG. 1. **a**, Selected temperature dependence of resistance for $Sb_2Te_3$ at various pressures showing a superconducting phase transition above 3K at 4.0 GPa. **b**, Magnetic-field dependence of the resistivity drop of $Sb_2Te_3$ at 6.7 GPa with an applied magnetic field *H* perpendicular to the *ab* plane of the single crystal. The dependence of $T_c$ on magnetic field *H* is shown in the inset of **b**.

FIG. 2. **a**, Pressure dependence of the superconducting transition temperature for $Sb_2Te_3$. **b**, Pressure-tuned changes on carrier density in $Sb_2Te_3$ at various temperatures. Solid and open circles indicate p-type and n-type characteristics, respectively.

FIG. 3. **a**, Synchrotron X-ray diffraction patterns of $Sb_2Te_3$ samples at selected pressures at 8 K. Arrows indicated the appearance of the diffractions peaks from high pressure phase. It is evident that the ambient pressure phase is stable at least up to 12 GPa, which shows that the pressure-induced superconductivity observed at the low-pressure range indeed comes from the ambient phase. **b**, Pressure dependence of the lattice parameters for the ambient pressure phase of $Sb_2Te_3$.

FIG. 4. Superconducting phase diagram of $Sb_2Te_3$ single crystals as a function of pressure. The green and yellow spheres in the α and β phases represent Sb and Te atoms, respectively, whereas the mixed color spheres in the γ and δ phases indicate that Sb and Te atoms are disordered and randomly occupied the lattice sites. Circles with various colors indicate the superconducting behaviors at different pressure phases. The superconductivity observed in the ambient phase of $Sb_2Te_3$ is labeled as TSC to indicate its topological nature.

Fig. 5. Bulk (upper panels) and (111) surface states (lower panels) of $Sb_2Te_3$ under 0 GPa (left panels) and 6.9 GPa (right panels).

FIG. 1(a)

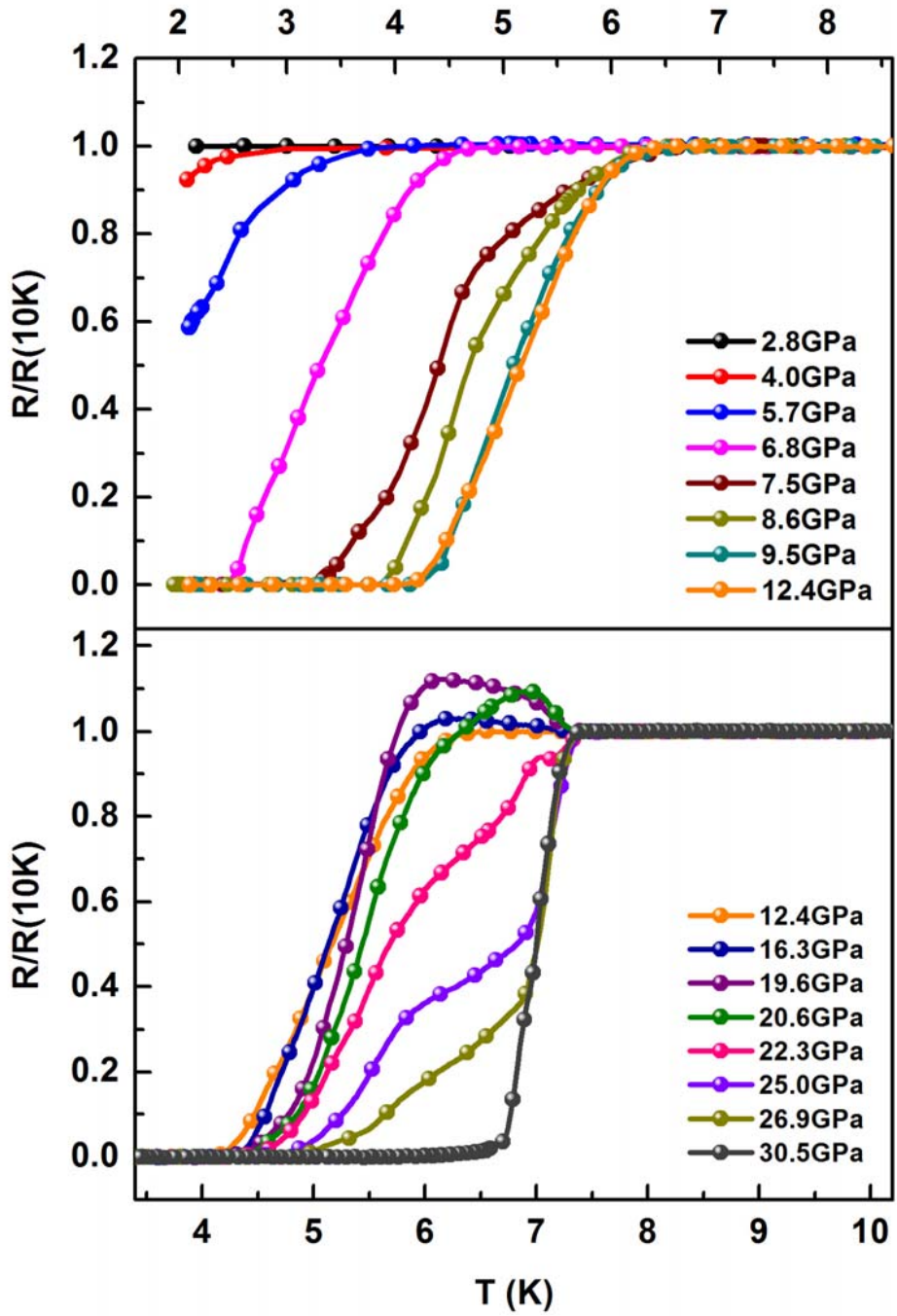

FIG. 1(b)

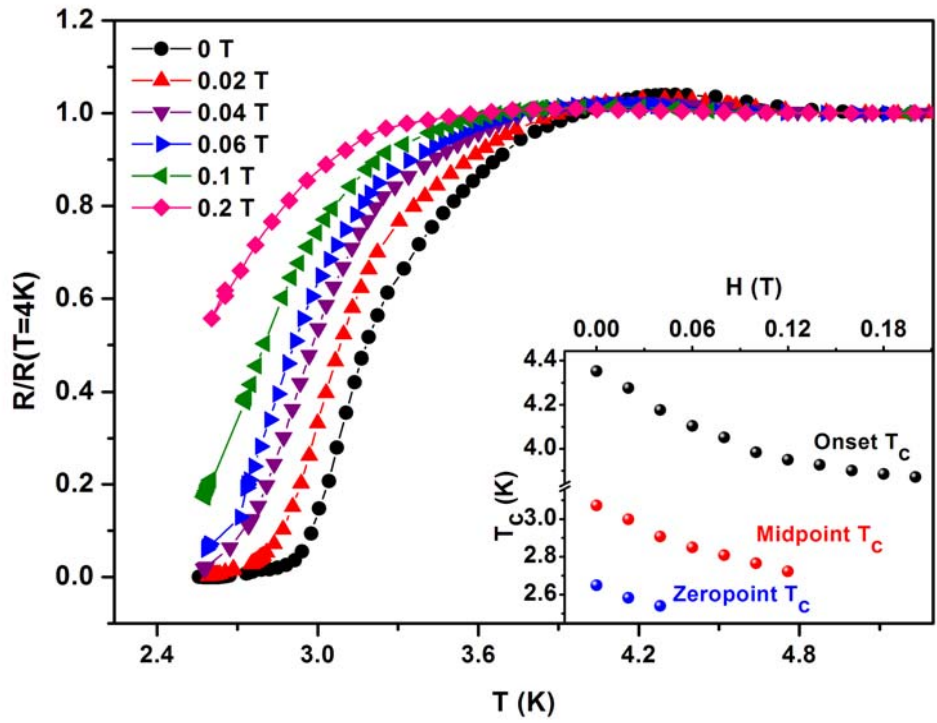

FIG. 2

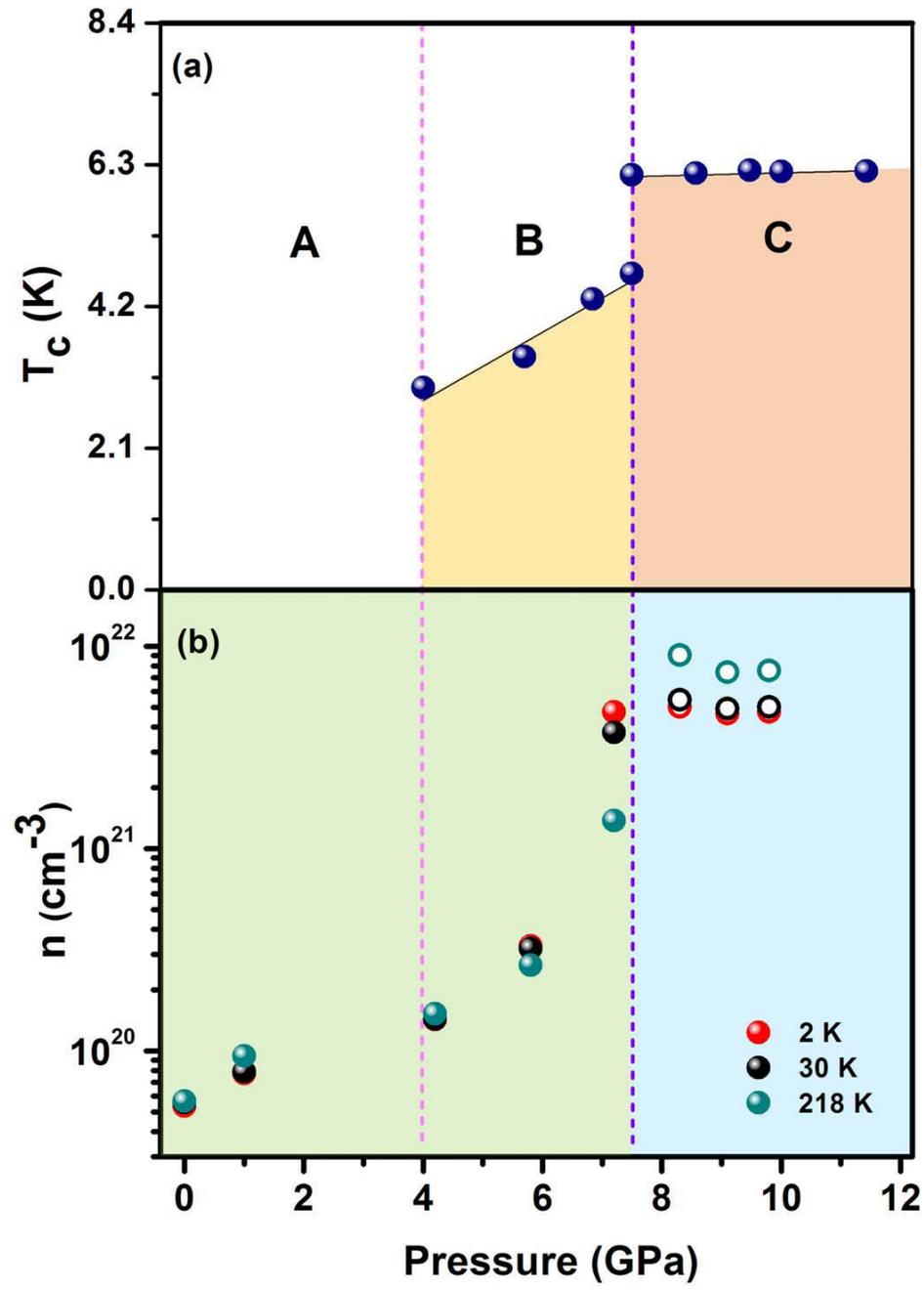

FIG. 3:

(a)

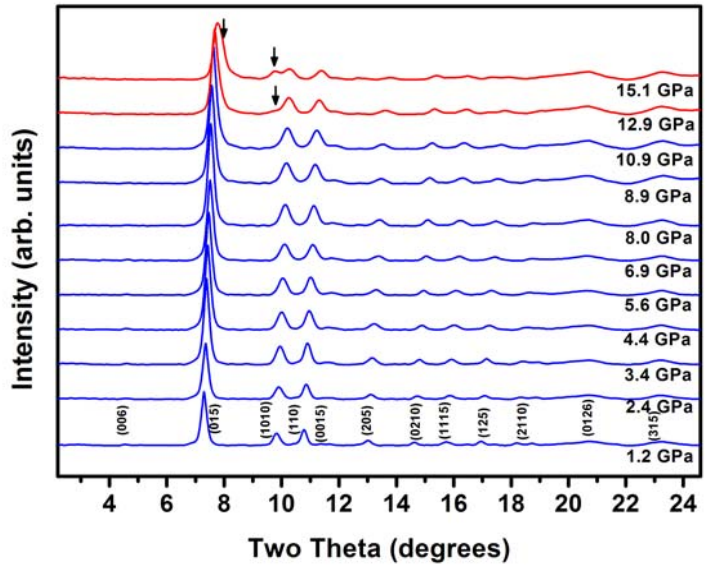

(b)

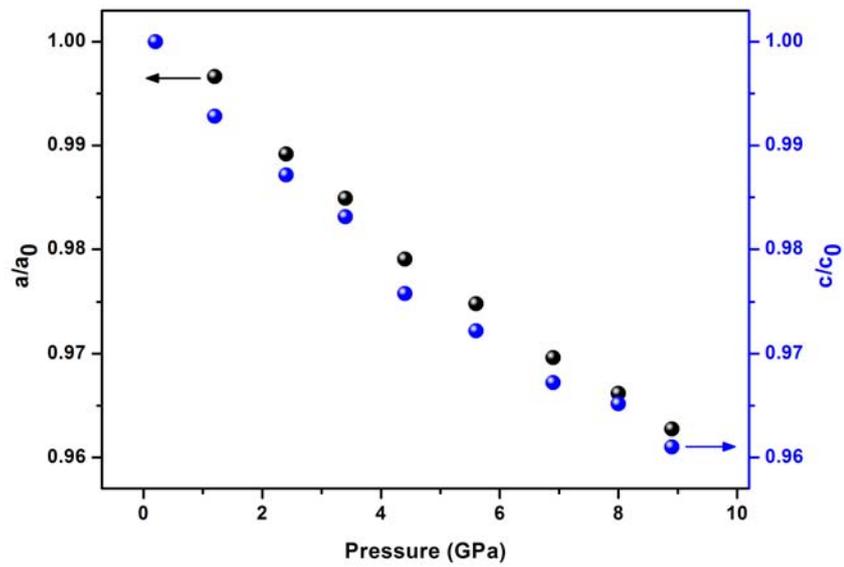

FIG.4

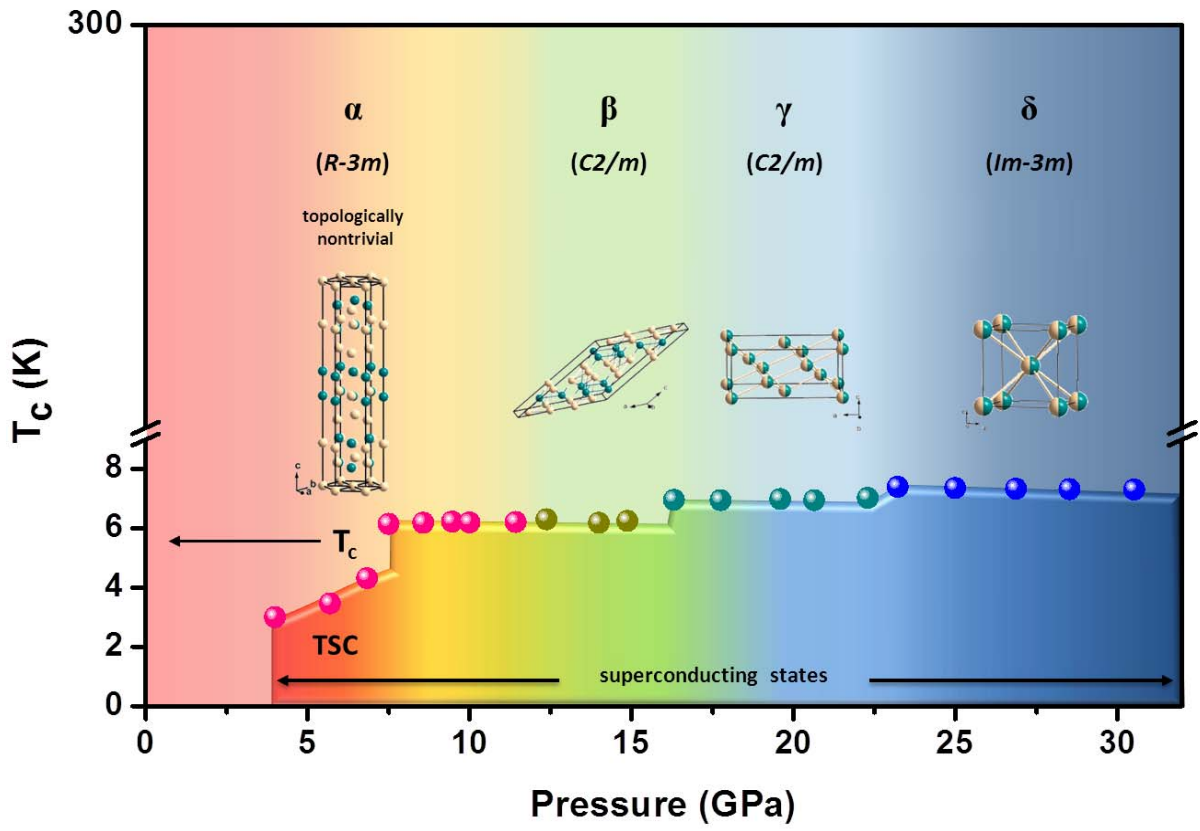

FIG.5

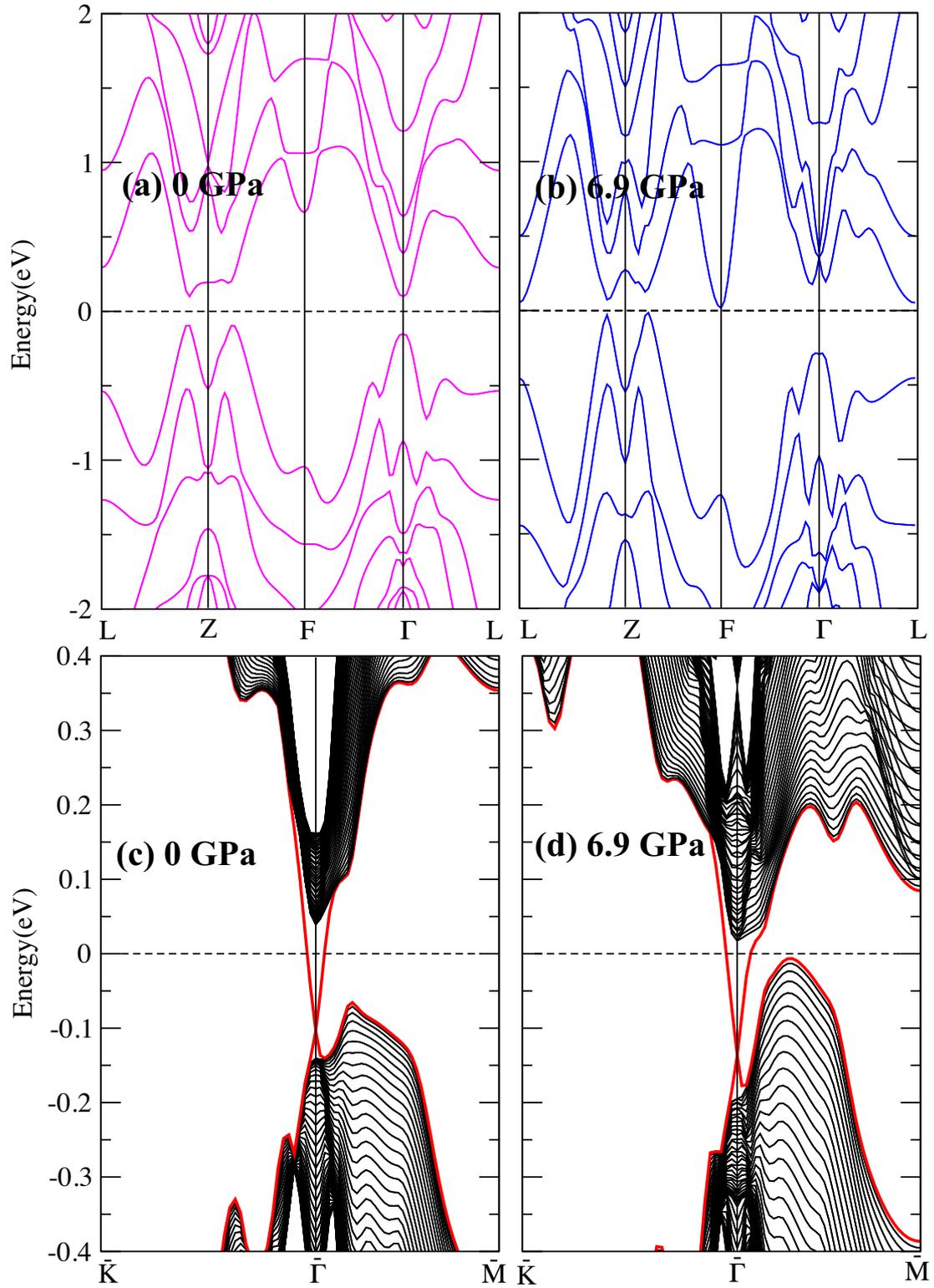